\documentstyle[eqsecnum,prl,aps]{revtex}
\begin{document}

\input{epsf.sty}

\draft
\def\now{\count100=\time \divide\count100 by 60 \count101=\time 
\count102=\count100 \multiply\count102 by 60 \advance\count101 by 
-\count102 \number\count100:\ifnum\count101<10 0\fi\number\count101}

\def\strike#1{[{\tt #1}]}
\def\add#1{\{{\bf #1}\}}
\preprint{\scriptsize Version: \today\ \ \now}
\title{Unique Signature of Dark Matter in Ancient Mica}
\author{Daniel P. Snowden-Ifft}
\address{Department of Physics, Amherst College, Amherst, 
Massachusetts 01002}
\author{Andrew J. Westphal}
\address{Department of Physics,
University of California at Berkeley, Berkeley, California  94720
}\maketitle
\begin{abstract}
Mica can store (for $> 1$ Gy) etchable tracks caused by atoms recoiling 
from WIMPs.  Because a 
background from fission neutrons will eventually limit this technique, a 
unique signature 
for WIMPs in ancient mica is needed.  Our motion around the center of the 
Galaxy causes WIMPs, unlike neutrons, to enter the mica from a preferred 
direction on the sky.  Mica is a directional detector and despite the 
complex 
rotations that natural mica crystals make with respect to this WIMP ``wind,'' there 
is a substantial dependence of etch pit density on present day mica 
orientation.
\end{abstract}
\noindent

Decades of research have established that at least 90\% of the mass of our 
universe is not emitting any light \cite{dark-matter-review,dark-matter-books}.  In recent years the nature 
of this dark matter has come under considerable experimental and 
theoretical scrutiny.  As a result of this scrutiny the list of potential 
candidates has dwindled from dozens a decade ago to only a few today 
\cite{dark-matter-books,dark-matter-detectors}.  Among the surviving 
candidates is the Weakly Interacting Massive Particle (WIMP).  From a 
theoretical standpoint WIMPs enjoy a great deal of support.  They fit 
well into our understanding of structure formation in the universe 
\cite{structure} and into many supersymmetric theories of particle physics 
\cite{supersymmetry}.  
Because of this theoretical support an increasing number of 
experimentalists employing a variety of techniques have set out to find 
these particles.  One such technique consists of looking for evidence of 
recoils from WIMPs in ancient mica\cite{snowden-ifft}. This technique 
exploits the 
fact that atoms recoiling from interactions with WIMPs
(``WIMP recoils'') would produce permanent chemical changes in the 
mica 
which later would produce etch pits when immersed in hydrofluoric acid.  
Stringent limits on WIMP cross sections as a function of mass were 
obtained \cite{snowden-ifft} 
by scanning only a small area, due to the fact that ancient mica 
integrates this signal for order 1 Gyr.  
Unfortunately other particles, e.g. fission neutrons, can also 
cause recoils which produce etch pits very similar to the etch pits we
expect from WIMP recoils.  Eventually this background will limit 
the 
technique and ultimately preclude the possibility of actually 
discovering 
WIMPs using this method.  A signature for WIMPs is needed.  
In this Letter we discuss a possible 
signature and calculate its magnitude.

The signature arises from the fact that our motion around the center 
of the Galaxy through a stationary WIMP halo causes a WIMP ``wind'' to 
arrive at Earth preferentially from one direction, the so-called
ram direction.  The ancient mica technique 
can exploit this asymmetry because the mica is itself asymmetric.
Mica is a layered crystalline mineral which is easily cleaved along the 
$\langle001\rangle$ 
crystallographic  plane.  In order to distinguish WIMP recoils from the 
large population of etch pits produced by alpha-decay in the $^{238}$U and 
$^{232}$Th decay chains, we require recoils to cross 
the cleavage plane \cite{snowden-ifft}.  In order to quantify the 
asymetry of the mica we define, 

\begin{equation}
s = {N_{\rm max} - N_{\rm min} \over N_{\rm max} + N_{\rm min}}  
\label{s-first}
\end{equation}
to be the signal contrast where N is the number of matched etched pits 
created by WIMP recoils and the subscripts, min and max, refer to the 
maximum and minimum of N over all possible mica orientations.  The signal 
contrast is the maximum assymetry possible.

To determine the magnitude of this parameter for a WIMP halo, we 
performed the following calculation.  Starting with the recoil spectrum\cite{spergel},
energy and angle with respect to  
the ram direction $\alpha$, and using $v_{\rm rms}
= 261\ {\rm km\ sec}^{-1}$ and $v_{\rm sun} = 220\ {\rm km\ sec}^{-1}$ 
with a cutoff at $v_{\rm c} = 640\ {\rm km\ sec}^{-1}$, we 
randomly generated recoils of the constituent atoms of mica.
Ranges and stopping powers of the recoils were calculated with the TRIM92 
computer code \cite{TRIM}.  As discussed  by 
Snowden-Ifft and Chan \cite{chan}, not every ion which crosses the cleavage 
plane produces an etch pit.  The appearance of an etch pit is determined 
by a stochastic process which depends on the type of stopping power (nuclear 
or electronic, characterized by two constants $k_n$ and $k_e$), the magnitude 
of the stopping power and the angle the recoil makes with the $\langle001\rangle$ 
plane.  The angular dependence of the etching model has been well tested \cite{chan} and it is 
found that ions normally incident on the cleavage plane are less likely 
to create an etch pit than those crossing at grazing angles.  Each recoil that 
crossed the cleavage plane 
was etched according to this etching model \cite{chan} with $k_e = 
0.8\times 10^{-5} 
({\rm MeV/g/cm}^2)^{-1}$, 
$k_n = 2.0\times 10^{-5} ({\rm MeV/g/cm}^2)^{-1}$ and $G = 
26$ nm.  A coherence term 
\cite{ahlen} was included in the calculation.  
This procedure was carried out for a variety of WIMP masses $m$ at a 
variety of angles $\alpha$.
We find that the rate of accumulation of matched etched 
pits with depths of 2 nm or greater can be adequately characterized
by 

\begin{equation}
	dN(\alpha)/dt = c_0(m) + c_1(m)|\cos\alpha|.
\label{n-eqn}
\end{equation}
$c_1$ was positive for all WIMP masses considered, so
mica oriented at $\alpha=0^\circ$ has a higher rate of accumulation 
than mica oriented at $\alpha=90^\circ$.  Two pieces of mica 
constantly held at $\alpha=0^\circ$ and $\alpha=90^\circ$ 
will therefore have the largest possible signal contrast, $\omega$, given by
  
\begin{equation}
\omega(m) \equiv c_1(m) / ( 2c_0(m) + c_1(m) ).
\label{smax-eqn}
\end{equation}
In Table \ref{table1} we show $\omega$ for several values of WIMP mass.  The lower the 
mass of the WIMP the larger $\omega$ although for very low mass WIMPs the 
expected density decreases rapidly.

No piece of mica has pointed in the same direction over geologic time, so 
to compute the WIMP track density as a function of contemporary orientation 
and 
age we require a detailed history of an ancient mica's orientation with 
respect to the ram direction.  This history is complex, but quantifiable.
Several rotations contribute to it; we discuss each below 
in order of increasing timescale.  The diurnal rotation of the 
vector normal to the $\langle001\rangle$ mica plane, the mica normal 
vector, about the celestial pole has the shortest timescale: 24 hours.  
Over timescales much larger than this 
the track density will be a function only of the declination of the 
mica normal vector, or the {\it mica normal declination}.
We assume that the interaction length
of WIMPs is much larger than the radius of the Earth.
The motion of the celestial pole with respect to the ecliptic pole is 
comprised of two general rotations:  nutation and precession.
Nutation has an amplitude of less than 10 arcseconds\cite{nutation}, 
so can be neglected.  
Precession is the smooth rotation of the celestial pole about the ecliptic pole with
fixed angle -- the obliquity -- between the two.  
The obliquities of most of the inner planets have
undergone large and chaotic changes on timescales of order 1-10 My 
\cite{laskar-robutel} since solar system formation, with excursions 
from $\sim 0^\circ$ to $> 90^\circ$.  Fortunately, 
both for this calculation and for the climatic stability of the earth, the 
presence of the moon has had a dramatic stabilizing effect, so that the 
earth's obliquity has varied from its mean value of $23.3^\circ$ by only 
$1.3^\circ$ since the moon was captured\cite{laskar-joutel-robutel}.

The slow orbital motion of the solar system around the center of the 
galaxy has the longest timescale of the celestial rotations.
This motion has the effect of slowly rotating the ram direction 
within the galactic plane.
We assumed that the solar system describes a circular orbit about the galactic
center, with a velocity of 220
km sec$^{-1}$\cite{galaxy-rotation} and a period of 224 My\cite{reid}.

In addition to these celestial rotations, tectonic drift, with a 
timescale between 10-100 My, also affects a mica crystal's orientation with 
respect to the ram direction.
It is not necessary 
to know the absolute longitude of the body, but a history of the latitude 
and orientation of the body is required.  Paleomagnetic data directly provide this history. 
In a 
simplified description of the technique, the direction of the terrestrial 
magnetic field is recorded by the magnetization direction of a magnetic 
mineral as it cools below the so called ``blocking''
temperature.  The local magnetic field direction at that time can thus be 
determined by a careful measurement of the magnetization direction of the 
mineral, and this determines 
the position of the terrestrial magnetic pole with respect to the sample.  
The age of the mineral is typically 
derived from the fossil record in the same stratum as the mineral for Phanerozoic
(younger than 590 My) samples, and by radioisotope or fission-track dating for 
pre-Cambrian (older than 590 My) samples.
In practice, there are many complications, but the technique is well-developed, 
and the position for an ancient magnetic pole can be determined with an accuracy of $\sim 5^\circ$.   The determination of a 
mean paleomagnetic pole is equivalent to a determination of the
geographic pole position relative to the sample\cite{mcelhinny}.  
The discrepancy between the mean paleomagnetic pole and the 
modern geographic pole is attributed to tectonic drift, yielding the latitude 
and orientation of the body containing the sample.

A large body of time sequences of paleomagnetic pole measurements
 --- generally known as Apparent Polar Wander Paths (APWPs) --- have been 
compiled by 
several authors (e.g., \cite{tarling,vandervoo}) with the general
goal of the reconstruction of ancient geography.  This 
dataset is ideal for our purposes.
The best-determined APWPs have been done for the major continental cratons 
(bodies which
have been geologically stable and have undergone only rigid-body rotations 
over long geological times).  For each craton, we arbitrarily choose a 
point 
near the geographic center of the craton for the purpose of calculation.   
As a general statement, the reconstruction of APWPs
is best for Phanerozoic time  because the fossil 
record allows for accurate dating;
for pre-Cambrian times  APWPs are in general rather poorly 
known, although the situation
is improving rapidly particularly for the North American and European 
continents (e.g., \cite{dagrella-filho}).

We briefly summarize here the method used for signal integration.
The track density for a piece of mica of age $t$, zenith angle $\theta_0$ and
azimuth angle $\phi_0$ (true north corresponding to $\phi_0=0^\circ$, 
true east to $\phi_0=90^\circ$, etc.) can be calculated from Eq. \ref{n-eqn} to be

\begin{equation}
N(t,\theta_0,\phi_0) = \left[c_0(m) + c_1(m)\langle|\cos\alpha|\rangle(t,\theta_0,\phi_0)\right] t,
\label{n-integrated}
\end{equation}
where $\langle|\cos\alpha|\rangle(t,\theta_0,\phi_0)$ is the average value of $|\cos\alpha|$
over the age of the mica sample.
For a given WIMP mass and mica age this function will achieve a maximum 
value, $N_{\rm max}$, in an orientation specified by $\theta_{\rm 
0,max}$ and $\phi_{\rm 0,max}$ and a minimum value, $N_{\rm min}$, for 
some other orientation specified by $\theta_{\rm 0,min}$ and $\phi_{\rm 
0,min}$.  Combining Eqs. \ref{s-first} and \ref{n-integrated} with the fact that 
$\langle|\cos\alpha|\rangle(t,\theta_{\rm 0,max},\phi_{\rm 0,max}) \sim
 \langle|\cos\alpha|\rangle(t,\theta_{\rm 0,min},\phi_{\rm 0,min}) \sim 0.5$ 
allows us to factor $s$ and write it as

\begin{equation}
 s(m,t) = \omega(m) \xi(t)
\label{s-factorization}
\end{equation}
where $\omega(m)$ (Eq. \ref{smax-eqn}) characterizes the asymmetry of the mica response and is 
independent of time, and

\begin{equation}
\xi(t) =  	\textstyle 
		\langle|\cos\alpha|\rangle(t,\theta_{\rm 0,max},\phi_{\rm 0,max}) -
		\langle|\cos\alpha|\rangle(t,\theta_{\rm 0,min},\phi_{\rm 0,min}).
\label{xi-eqn}
\end{equation}
characterizes the mica's history and is independent of mass.

\begin{figure}
\epsfxsize=2.5in
\epsfbox{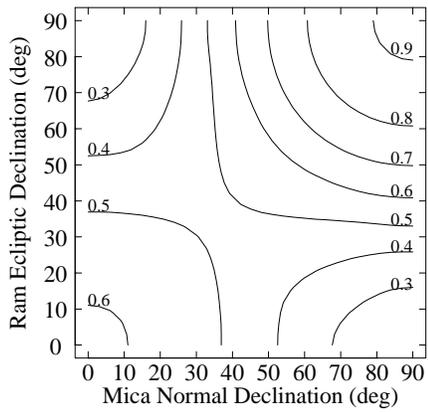}
\caption{Contour of values of $\langle|\cos\alpha|\rangle$,
averaged over times much longer than the precessional period, as a function
of Mica Normal Declination and Ram Ecliptic Declination.}
\label{contour}
\end{figure}

The track density integrated over timescales
longer than the precessional timescale ($\sim 26\,000$ y) but shorter than the timescale for
tectonic drift and galactic rotation ($\sim 10$ My) is
a function only of mica normal declination $\delta_{\rm mica}$ and of the ecliptic declination of the ram 
direction $\delta_{\rm ram}$,
since rapid diurnal rotation and polar precession have the effect of
averaging over celestial right ascension and ecliptic longitude. 
In Fig. \ref{contour}, we show $\langle|\cos\alpha|\rangle_{\rm int}(\delta_{\rm ram},\delta_{\rm mica})$,
the calculated average of $|\cos\alpha|$
over these intermediate timescales 
as a function of mica normal declination and ram ecliptic declination.

To calculate  $\langle|\cos\alpha|\rangle$ over geologic timescales,
we first determine the ram ecliptic declination $\delta_{\rm ram}(t^\prime)$
for a given short geological time interval $(t^\prime, t^\prime + 
dt^\prime)$.
We then determine the continental latitude and orientation using an 
linear interpolation on the sphere 
between the two paleomagnetic pole measurements closest in time to 
$t^\prime$.
The ancient values of mica normal declination $\delta_{\rm mica}(t^\prime)$ are 
determined for each contemporary mica sample
orientation  ($\theta_0$,$\phi_0$),  using the 
paleomagnetically determined rotations from the current position.   
Symbolically, the functional dependence of the calculations is summarized 
by:

\begin{equation}
 \langle|\cos\alpha|\rangle(t,\theta_0,\phi_0) = {1\over t}\int_0^t 
dt^\prime 
	\langle|\cos\alpha|\rangle_{\rm int}\left[\delta_{\rm ram}(t^\prime),\delta_{\rm mica}(t^\prime,\theta_0,\phi_0)\right]. 
\end{equation}
$\xi(t)$ in Eq. 6 may then be calculated using Eq. 7.
Although much of this integration can be in principle be done analytically, 
tectonic drift requires that the calculation be done numerically. 

\begin{figure}
\epsfxsize=2.5in
\epsfbox{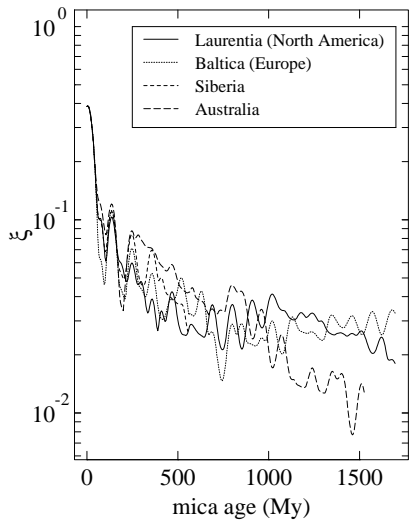}
\caption{Signal amplitude expressed as a fraction $\xi$ of the maximal signal 
amplitude $\omega$, plotted as a function of mica age for samples from 
four major continental cratons.}
\label{xifigure}
\end{figure}

In Fig. \ref{xifigure}, we show $\xi(t)$ as a function of mica age for  
several major continental cratons.  For this calculation, we used time 
intervals
of 5 My,  a $50 \times 50$ grid of diurnal and precessional rotations, and 
a $50 \times 50$ grid of mica orientations.
Because of the symmetry of the mica  ---
mica orientation is defined only up to a parity transformation ---
it is only necessary to define the signal over the upper hemisphere, ($0 
\le \theta \le 90^\circ$).  The periodicity due to galactic rotation 
is clearly visible.  It is desirable to choose locations and ages for 
which $\xi$ is large and, because of uncertainties in measuring 
the age of the 
mica, for which the signal amplitude and direction are not changing 
rapidly.  
We intend to address in a future paper the effect of
the uncertainties in our calculation, including
the uncertainty in the length of the galactic year ($\sim 20\%$), the
effect of uncertainties in mica ages, and so on.
For old mica the 
signal is smaller but the integration time is longer and for a 
given area of mica scanned the statistical significance of an observed 
signal turns out to be roughly independent of age.  The signal contrast for
50 GeV WIMPs ($\omega = 17$\%) in mica 500 My old ($\xi \sim 
6$\%) is of order one percent ($s \sim 1$\%).  This signal contrast is small
but it is ``state-of-the-art.''  No WIMP limits have been published with detectors
having a substantially larger signal contrast.

Finally we need to address the feasibility of such an experiment.  First 
in order to utilize this idea we will need to obtain several pieces of 
mica oriented in the proper directions.  Fortunately the large nearly 
perfect crystals required of such a search are almost always found in 
coarse grained rock formations known as pegmatites.  A typical pegmatite 
will contain many large mica crystals oriented in random directions\cite{klein}.  The 
existence of these structures virtually guarantees that we will be able 
to obtain mica of the proper quality having the same composition and same 
history and oriented in nearly the optimal directions.  Large 
concentrations of uranium in the pegmatite will need to be surveyed as 
the fast neutrons that emanate from them could potentially mimic our 
signal.  The second issue is the difficulty of detecting a $s \sim 1$\% signal.
If WIMPs exist at cross-sections just below our current limits we 
estimate that it will take a $\sim 4 \times 10^4$ improvement in scanned 
area to see a signal at the 3$\sigma$ level.  With an automated AFM an area equivalent 
to that of our first mica search can be scanned in a little less than 4 
hours.  An additional factor of 3 is gained by etching for twice as 
long.  As mentioned above the appearance of an etch pit is governed by a 
stochastic process.  As shown in \cite{chan}, a longer etching time 
results in more opportunities for an etch pit to form.  This effect has 
been confirmed experimentally\cite{unpublished}.  The $\sim 4 \times 10^4$ improvement can 
therefore be achieved in about 6 
years of machine time.  As discussed in [6] large crystals of mica will 
allow us to achieve this improvement without running into neutron 
background.  If a signal is observed it can be confirmed by similar 
experiments on other pegmatites.

We conclude by first stressing the two main points of this Letter.  
Because of the asymmetries inherent in etching latent tracks in ancient 
mica our detector can sense the direction of the WIMP ``wind.''  Moreover 
despite the complex rotations that mica crystals makes with respect to 
this 
WIMP ``wind,'' there is a substantial dependence of etch pit density on 
present day mica orientation.  This dependence can be used as a signature 
for WIMPs which in turn will allow us to detect WIMPs if they exist or 
rule them out more efficiently if they do not.

\section{Acknowledgments}

We are very grateful to Rob Van Der Voo, who kindly provided us with 
invaluable Pre-Cambrian APWPs.  We thank Paul Renne, Yudong He, and Buford Price 
for useful discussions, suggestions, and support.

\bibliographystyle{unsrt}

{\narrowtext
\begin{table}
\caption{ Calculated values of track accumulation rate $c_0$, angular coefficient of track accumulation
rate $c_1$, and signal contrast $\omega$  for
several values of WIMP mass.  The error bars arise from the Monte 
Carlo technique used to calculate these points.}
\begin{tabular}{|c|c|c|c|}
WIMP mass 	& $c_0$ 			& $c_1$ 			& $\omega$ \\
(GeV)		& (cm$^{-2}$ Gyr$^{-1}$)	& (cm$^{-2}$ Gyr$^{-1}$) 	&		\\
\tableline
\tableline
50 		& 2\,300\,000 $\pm$ 100\,000  		& 900\,000 $\pm$ 100\,000  		& 0.17 $\pm$ 0.04 \\
100 	& 2\,860\,000 $\pm$ 80\,000  		& 700\,000 $\pm$ 100\,000  		& 0.11 $\pm$ 0.03 \\
300 	& 1\,640\,000 $\pm$ 30\,000  		& 230\,000 $\pm$ 40\,000  		& 0.07 $\pm$ 0.02 \\
1000 	& 576\,000 $\pm$ 9\,000  		    & 80\,000 $\pm$ 10\,000  		& 0.06 $\pm$ 0.02 \\
10000 	& 63\,000 $\pm$ 1\,000  	     	& 6\,000 $\pm$ 2\,000  	    	& 0.04 $\pm$ 0.02 \\
\end{tabular}
\label{table1}
\end{table}
}

\end{document}